\documentclass{elsart}
\usepackage{amssymb}
\journal{Physics Letter B}
\begin{document}
\begin{frontmatter}
\title{Cosmic Strings and Quintessence}
\author[lu]{Yi-Shi Duan},
\author[lu]{Ji-Rong Ren\corauthref{cor}} and  \corauth[cor]{Corresponding author} \ead{renjr@lzu.edu.cn}
\author[lu]{Jie Yang}
\address[lu]{The Institute of Theoretic Physics, Lanzhou University 730000, P.R.China}

\begin{abstract}
\indent \hskip 16pt We present a new Lorentz gauge invariant
$U(1)$ topological field theory in Riemann-Cartan spacetime
manifold $U_{4}$. By virtue of the decomposition theory of $U(1)$
gauge potential and the $ \phi $--mapping topological current
theory, it is proved that the $U(1)$ complex scalar field $\phi
(x)$ can be looked upon as the order parameter field in our
Universe, and the set of zero points of $\phi (x)$ create the
cosmic strings as the spacetime defects in the early Universe. In
the standard cosmology this complex scalar order parameter field
possesses negative pressure, provides an accelerating expansion of
Universe and be able to explain the inflation in the early
Universe. Therefore this complex scalar field is not only the
order parameter field created the cosmic strings, but also
reasonably behaves as the quintessence, the dark energy.
\end{abstract}
\end{frontmatter}
\parskip -0mm
\section{Introduction}
\indent \hskip 16pt The topological spacetime defects such as
cosmic strings may have been found at phase transitions in the
early Universe, analogous to those examples found in some
condensed matter systems---vortex line in liquid helium, flux
tubes in type-II superconductors, or disclination lines in liquid
crystals\cite{duanzhang,dysz}. Such phase transitions in the early
Universe are usually viewed as being caused by the breaking of the
symmetry of a Higgs field occurring in a spacetime whose geometry
is Riemannian\cite{Baker}. However the Einstein-Cartan theory,
which is formulated in a Riemann-Cartan spacetime manifold $U_4$,
represents a viable gravity theory with torsion. So it is still an
open debate if the spacetime manifold is Riemann-Cartan or not.
Furthermore, the viewpoint that the early Universe is
Riemann-Cartan spacetime in which cosmic strings are treated as
line defects created directly from torsion\cite{sabb}, is very
attractive, but it is not very perfect\cite{duanzhang,dysz,dyj}.
At the same time the order parameter of the Universe is not very
clear yet. In this letter, by virtue of constructing a new Lorentz
gauge invariant $U(1)$\ topological field from torsion, we present
a theory of creating cosmic strings. Further this theory
rigorously unites the geometry and topology of gauge theories of
spacetime defects and is independent of concrete spacetime
manifold. It's also important to show that the order parameter
field $\phi =\phi ^1+i\phi ^2$\ which creating cosmic strings,
possesses negative pressure. Therefore the complex scalar order
parameter field has the property of quintessence, the dark energy.
\par \indent \hskip 16pt The
theoretical framework in the present letter includes three basic
aspects. In section 2, we present a new Lorentz gauge invariant $
U(1)$\ topological field theory in $U_4$. This general theory is
still valid in both special cases: the Riemann manifold and the
Weitzenb$\ddot{o}$ck manifold. In section 3, by virtue of the
decomposition theory of gauge potential and the $\phi$--mapping
topological current theory\cite{DuanFZ,duanmeng}, the cosmic
strings as the spacetime defects are created naturally from the
zero points set of the order parameter $\phi $\ and the length of
these strings are topologically quantized by winding number of
$\phi $--mapping with Planck length ($L_p=\sqrt{\frac{\hbar
G}{c^3}}$). In section 4, The nonzero part of the complex scalar
order parameter field $\phi $\ is shown as quintessence which
possesses negative pressure and cause inflation of the early
Universe.

\section{ A New Topological Invariant in Riemann-Cartan Manifold}

\indent \hskip 16pt It is well known that in a 4-dimensional
Riemann-Cartan spacetime manifold $U_4$, the torsion $2$-form is
defined by
\begin{equation}
T^a=De^a,  \label{tors}
\end{equation}
where $e^a$\ is the vierbein $1$-form with Lorentz group index $a$
and
\begin{equation}
e^a=e_\mu ^adx^\mu .  \label{vd}
\end{equation}
The vierbein $e_\mu ^a$\ is related with the metric tensor by
\begin{equation}
g_{\mu \nu }=e_\mu ^ae_\nu ^a{\ \ ,\ \ }\mu ,\nu ,a=0,1,2,3.
\label{me}
\end{equation}
$D$\ is the covariant differential in Lorentz gauge theory
\begin{equation}
D=D_\mu dx^\mu \quad ,\quad D_\mu =\partial _\mu -\omega _\mu ,
\label{der}
\end{equation}
\begin{equation}
\omega _\mu =\frac{1}{2}\omega _\mu ^{ab}I_{ab},  \label{ome}
\end{equation}
where $\omega _\mu ^{ab}$\ is the spin connection and $I_{ab}$\ is
the generator of lorentz group $SO(1,3)$. The torsion $2$-form can
be expressed explicitly by
\begin{equation}
T^a=\frac{1}{2}T_{\mu \nu }^adx^\mu \wedge dx^\nu ,  \label{tors0}
\end{equation}
in which the torsion tensor is
\begin{equation}
T_{\mu \nu }^a=D_\mu e_v^a-D_ve_\mu ^a.  \label{tors1}
\end{equation}
\indent \hskip 16pt  As in ref.\cite{dysz}, by analogy with the 't
Hooft's viewpoint of $SU(2)$\ gauge theory of
monopole\cite{Hooft}, using vierbein $1$-form and torsion
$2$-form, we define a Lorentz gauge invariant antisymmetrical
$2$-form $K$\
\begin{equation}
K=T^au^a-e^a\wedge Du^a,  \label{kf0}
\end{equation}
where $u^a$\ is a\ $SO(1,3)$\ unit vector field. It's easy to see
that $K$\ can be expressed explicitly as
\begin{equation}
K=\frac{1}{2}K_{\mu \nu }dx^\mu \wedge dx^\nu ,  \label{kf1}
\end{equation}
and $K_{\mu \nu }$ is an antisymmetric tensor
\begin{equation}
K_{\mu \nu }=T_{\mu \nu }^au^a-\left( e_\mu ^aD_\nu u^a-e_\nu
^aD_\mu u^a\right) .  \label{nt}
\end{equation}
\indent \hskip 16pt  Substituting torsion tensor formula
(\ref{tors1}) into (\ref{nt}), we can find a simplified form for
$K_{\mu \nu }$
\begin{equation}
K_{\mu \nu }=\partial _\mu \left( e_\nu ^au^a\right) -\partial
_\nu \left( e_\mu ^au^a\right) ,  \label{kmn}
\end{equation}
which shows that the dimensionless vector $e_\nu ^au^a$ in
(\ref{kmn}) can be looked upon as a $U(1)$\ gauge potential. Since
the usual $U(1)$\ gauge potential $a_\mu $\ in covariant
derivative $D_\mu =\frac{\partial}{\partial x^\mu }-ia_\mu$ must
has dimension $[ L ] ^{-1}$, we can define a $U(1)$ gauge
potential
\begin{equation}
A_\mu =\frac{1}{L_P}e_\mu ^au^a,  \label{gp}
\end{equation}
where $L_P$ is the Plank length introduced to satisfy the
dimension of $ U(1)$ gauge potential. The antisymmetric tensor
$K_{\mu \nu }$ exactly behaves as a $U(1)$ gauge field tensor
\begin{equation}
K_{\mu \nu }=L_P\left( \partial _\mu A_\nu -\partial _\nu A_\mu
\right) . \label{ugf}
\end{equation}
And the $U(1)$ gauge transformation of $A_\mu $ is
\begin{equation}
A_\mu ^{\prime }=A_\mu +\partial _\mu \alpha ,  \label{gt}
\end{equation}
where $\alpha (x)$ is an arbitrary scalar function.
\par \indent \hskip 16pt
Since for a given Riemann-Cartan manifold the vierbein $ e_\mu ^a$
is fixed, then the above $U(1)$ gauge transformation (\ref{gt})
can be only explained as the change of the unit vector
$u^a\rightarrow u^{a^{\prime }}$, therefore
\begin{equation}
e_\mu ^au^{a^{\prime }}=e_\mu ^au^a+\partial _\mu \alpha ,
\label{gt1}
\end{equation}
which gives
\begin{equation}
u^{a^{\prime }}=u^a+e^{\mu a}\partial _\mu \alpha ,  \label{uvp}
\end{equation}
where $e^{\mu a}$ is the inverse vierbein
\begin{equation}
e_\mu ^ae^{\mu b}=\delta ^{ab}.  \label{ivv}
\end{equation}
\indent \hskip 16pt  It must be pointed out that the $SO(1,3)$
gauge invariant antisymmetry tensor $K_{\mu \nu }$ can also be
defined as\cite {duanzhang,dysz}
\begin{equation}
K_{\mu \nu }=T_{\mu \nu }^au^a,  \label{cc}
\end{equation}
with a parallel $SO(1,3)$ unit vector field $u^a$ satisfied
\begin{equation}
Du^a=0\quad ,\quad a=0,1,2,3.  \label{uvf}
\end{equation}
\indent \hskip 16pt  Therefore we have presented a new Lorentz
($SO(1,3)$) gauge invariant $U(1)$ field theory in $U_4$. In this
new $U(1)$ gauge theory the corresponding first Chern class in
$U_4$ is a topological invariant with dimension $[L]$ and is
expressed as
\begin{equation}
L=\frac{1}{2\pi }\int_\Sigma K,  \label{fc0}
\end{equation}
where $\Sigma $ is an arbitrary 2-dimension surface in $U_4$.
Substituting ( \ref{kf1}) and (\ref{nt}) into (\ref{fc0}), it is
easy to see that the complete expression for our new $U(1)$
topological invariant in $U_4$ is
\begin{equation}
L=\frac{1}{2\pi }\int_\Sigma \left\{ \frac{1}{2}\left[ T_{\mu \nu
}^au^a-\left( e_\mu ^aD_\nu u^a-e_\nu ^aD_\mu u^a\right) \right]
\right\} dx^\mu \wedge dx^\nu . \label{fc1}
\end{equation}
\indent \hskip 16pt  In the last years, it is still an open debate
whether the spacetime of the early Universe is Riemann-Cartan
manifold or not. Here we should point out that in torsion free
case
\begin{equation}
T_{\mu \nu }^a=0\quad ,\quad D_\mu e_v^a=D_ve_\mu ^a,  \label{tfc}
\end{equation}
and from (\ref{fc1}) we can find that the formula (\ref{ugf}) and
(\ref{fc0} ) still holds true. Therefore the theory mentioned
above may be considered as a universal $U(1)$ topological
invariant theory in Riemann-Cartan spacetime manifold $U_4$ and
Riemann spacetime manifold $V_4$.

\section{Cosmic Strings}

\indent \hskip 16pt In order to introduce the decomposition theory
of $U(1)$
 gauge potential\cite{DuanFZ} in $U_4$, let the complex scalar
\begin{equation}
\phi =\phi ^1+i\phi ^2  \label{phi}
\end{equation}
be an ordered parameter field in our Universe and its covariant
derivative corresponding gauge potentials $A_\mu $ (\ref{gp})
\begin{equation}
D_\mu \phi =\partial _\mu \phi -iA_\mu \phi .  \label{gpcd}
\end{equation}
By means of the decomposition theory of U(1) gauge potential
proposed by one of authors (Duan)\cite{duanmeng}, the gauge
potential (\ref{gp}) can be decomposed\cite{DuanJiang} as
following
\begin{equation}
A_\mu =\epsilon _{AB}n^A\partial _\mu n^B  \label{gpd}
\end{equation}
where $n^A$ ($A=1,2$) is 2-dimensional unit vector fields
\begin{equation}
n^A=\frac{\phi ^A}{\|\phi \|}\quad ,\quad n^An^A=1\quad ,\quad
A=1,2.
\end{equation}
 Substituting (\ref{gpd}) into (\ref{ugf}), we have
\begin{equation}
K_{\mu \nu }=L_p\epsilon _{AB}(\partial _\mu n^A\partial _\nu
n^B-\partial _\nu n^A\partial _\mu n^B),  \label{git}
\end{equation}
and the corresponding new topological invariant (\ref{fc0}) is
denoted by
\begin{equation}
L=\frac{L_p}{2\pi }\int_\Sigma \epsilon _{AB}\partial _\mu
n^A\partial _\nu n^Bdx^\mu \wedge dx^\nu .  \label{1c}
\end{equation}
By making use of the $\phi $-mapping topological current
theory\cite {duanmeng} and the Green function equation in $\phi
$-space
\begin{equation}
\bigtriangleup _{{\bf \phi }}\ln \| \phi \| =2\pi \delta ^2({\bf
\phi })\quad ,\quad \bigtriangleup _{{\bf \phi }}=\frac{\partial
^2}{\partial \phi ^A\partial \phi ^A},
\end{equation}
we can find
\begin{equation}
L=L_p\int_\Sigma \delta ^2(\phi )D_{\mu \nu }\left( \frac{\phi}{x}
\right) dx^\mu \wedge dx^\nu ,  \label{L}
\end{equation}
where $D_{\mu \nu }\left( \frac{\phi }{x}  \right) =$ $\epsilon
_{ab}\partial _\mu ^A\phi \partial _\nu \phi ^B$ is the Jacobian
tensor. The 2-dimension surface $\Sigma $ can be parameterized by
\begin{equation}
x^\mu =x^\mu \left( u^1,u^2\right) ,  \label{sf}
\end{equation}
and the integral $L$ (\ref{L}) is expressed as
\begin{equation}
L=L_p\int_\Sigma \delta ^2(\phi )D\left( \frac{\phi}{u}  \right)
\sqrt{g_u}d^2u. \label{iL}
\end{equation}
It is obviously that the above integral $L$ (\ref{iL}) is non-zero
only when
\begin{equation}
\phi ^A(\stackrel{\rightarrow }{x},t)=0\quad \quad ,\quad \quad
A=1,2. \label{Phi0}
\end{equation}
Suppose that the order parameter field $\phi ^A(x)$ ($A=1,2$)
possesses $ \ell $ zeroes, according to the implicit function
theorem\cite{goursat1}, when the zeroes are regular points of
$\phi $-mapping at which the rank of Jacobian matrix $[\partial
_u\phi ^A]$ is 2, the $i$-th solution of $\phi
^A(\stackrel{\rightarrow }{x},t)=0$  ($A=1,2$) can be expressed by
\begin{equation}
S_i:x^\mu =z_i^\mu \left( s,t\right) \quad ,\quad i=1,2,...,\ell .
\label{S}
\end{equation}
We see that each solution of (\ref{S}) is a world sheet of a
string with string parameter $s$ and time $t$.
\par \indent \hskip 16pt  Using the expanding formula of $\delta
^2(\phi )$ in the $\phi $-mapping topological current
theory\cite{fudz} we have
\begin{equation}
L=L_p\int_\Sigma \sum_{i=1}^\ell W_i\delta ^2(u-z_i)d^2u,
\label{Lp}
\end{equation}
where $z_i (i=1,2,...,\ell )$ are the intersection points of
strings $S_i$
 with surface $\Sigma $ and $W_i$ is winding number of $\phi ^A(x)$ ($
A=1,2$) around string $S_i$ on $\Sigma $ at $z_i$. Then we have
\begin{equation}
L=\sum_{i=1}^\ell L_i\quad ,\quad L_i=L_pW_i.  \label{St}
\end{equation}
\indent \hskip 16pt  This is our $\phi $-mapping topological
current theory of cosmic strings in $U_4$ and $V_4$, which shows
that the cosmic strings are created from the zeros of complex
scalar ordered parameter field $\phi ^A(x)$ ($ A=1,2$) and the
length $L_i$ of each string is topologically quantized by winding
numbers with Planck length $L_p$ .

\section{Complex Scalar Quintessence}

\indent \hskip 16pt Recent observation of type Ia supernova(SNe
Ia) revealed that the expansion of the Universe is
accelerating\cite{Perlmutter}. This means the pressure of Universe
is negative. The result of BOOMERanG supports a flat Universe
($k=0$) which leads to that the total density of the Universe is
equal to the so-called critical density\cite {Bernardis}. The
properties of these observed have important implication for
cosmology that the Universe consists of 1/200th bright stars,
$1/3rds$ dark matter and $ 2/3rds$ ``dark energy'' with a negative
pressure as $p<-\rho c^2/3$ \cite{turner1}. The latest
observations of distant quasars distorted by massive invisible
objects have provided fresh evidence that the Universe is mostly
made up of mysterious ``dark energy''\cite{cb}. In the light of
these observations, some cosmologists called the exotic dark
energy as quintessence, a slowly evolving scalar or complex
scalar\cite{GuHwang}. The main goal of this section is to show
that the complex scalar field acting as the order parameter which
creates the cosmic strings can also lead to both negative pressure
and acceleration of Universe. Therefore the complex scalar order
parameter field introduced in the last section can play an
important role of quintessence, so-called dark energy.
\par \indent \hskip 16pt  In the standard cosmology our Universe is
described by the Friedman-Robertson-Walker (FRW) metric:
\begin{equation}
ds^2=c^2dt^2-a^2(t)(\frac{dr^2}{1-kr^2}+r^2d\theta ^2+r^2\sin
^2\theta d\phi ^2),  \label{frw}
\end{equation}
here $a(t)$ is the cosmic scale factor, $k$ is the signature of
curvature. The cosmology dynamic equations are well-known as
following
\begin{equation}
\frac{\stackrel{..}{a}}{a}=-\frac{4\pi G}{3c^2}(\rho c^2+3p),
\label{cde1}
\end{equation}
\begin{equation}
\stackrel{\cdot }{a}^2+kc^2=\frac{8\pi G}3\rho a^2.  \label{cde2}
\end{equation}
The eq.(\ref{cde2}) leads to
\begin{equation}
k=\frac{8\pi G}{3c^2}\left( \rho -\rho _c\right) a^2,  \label{k}
\end{equation}
where $\rho _c$ is the critical density which is related to the
hubble constant.
\par \indent \hskip 16pt  Suppose that the
Lagrangian density for the complex scalar order parameter field is
\begin{equation}
{\mathcal{L}} = \frac{1}{2}g^{\mu \nu }\partial _\mu \phi
^A\partial _\nu \phi ^A-V(\phi ), \label{lag}
\end{equation}
where $\phi ^A$ $(A=1,2)$ are real and imaginary parts of the
complex scalar order parameter field $\phi =\phi ^1+i\phi ^2$ and
$V(\phi )$ is a potential function of $\phi ^{*}\phi =\phi ^A\phi
^A$. In the last section, using the $\phi $-mapping theory we have
shown that the singularities of $ \delta (\phi )$ can create the
cosmic strings. As in the case of scalar
field\cite{Ratra,Wetterich,Coble,Turner}, the perfect fluid model
in standard cosmology for Lagrangian (\ref{lag}) gives
\begin{equation}
\rho _\phi c^2=\frac{1}{2}\stackrel{.}{\phi ^A}\stackrel{.}{\phi
^A}+\frac{1}{ 2a^2}\nabla \phi ^A\cdot \nabla \phi ^A+V(\phi ),
\label{ed0}
\end{equation}
\begin{equation}
p_\phi =\frac{1}{2}\stackrel{.}{\phi ^A}\stackrel{.}{\phi
^A}-\frac{1}{6a^2} \nabla \phi ^A\cdot \nabla \phi ^A-V(\phi ).
\label{pre0}
\end{equation}
As we know the scale factor $a(t)$ is very large in the present
Universe, and the above two equations become
\begin{equation}
\rho _\phi c^2=\frac{1}{2}\stackrel{.}{\phi ^A}\stackrel{.}{\phi
^A}+V(\phi ), \label{ed}
\end{equation}
\begin{equation}
p_\phi =\frac{1}{2}\stackrel{.}{\phi ^A}\stackrel{.}{\phi
^A}-V(\phi ). \label{pre}
\end{equation}
\indent \hskip 16pt  As usual $\phi ^A$ $(A=1,2)$ is considered as
slowly evolving\cite{Caldwell}, i.e. approximately static
\begin{equation}
\stackrel{.}{\phi ^A}\approx 0,  \label{sp}
\end{equation}
then we have
\begin{equation}
\rho _\phi c^2\approx +V(\phi )\quad ,\quad p_\phi \approx -V(\phi
), \label{eq}
\end{equation}
and
\begin{equation}
p_\phi \approx -\rho _\phi c^2.  \label{pr}
\end{equation}
Thus, the complex scalar order parameter possesses a negative
pressure $ p_\phi <-1/3\rho _\phi c^2,$ which leads to an
accelerating expansion Universe by eq.(\ref{cde1}).
\par \indent \hskip 16pt  The observations of the power spectrum of
cosmic microwave background (CMB) indicate that the Universe is a
spatially flat Universe( $k=0$) and from (\ref{k}) we have
\begin{equation}
\rho =\rho _c,  \label{fcd}
\end{equation}
i.e. total density of the Universe is equal to the critical
density $\rho _c$. As mentioned in the beginning of this section
the recent observations implicate that the quintessence is enough
to explain the unknown $2/3rds$ of critical density. Therefore the
complex scalar order parameter field with both negative pressure
and creating the cosmic strings is a reasonable candidate for the
quintessence(dark energy).
\par \indent \hskip 16pt  It must be
pointed out that from (\ref{pr}) and the equation of motion of the
perfect fluid with FRW metric, we can find a inflation solution of
the Universe
\begin{equation}
a(t)\propto e^{\chi t},  \label{ss}
\end{equation}
where
\begin{equation}
\chi =\sqrt{\frac{8\pi G}3\rho _c}\sim 10^{34}\sec ^{-1}.
\label{ka}
\end{equation}
This shows that in a very small time interval
\begin{equation}
\triangle t=t-t_0\approx 10^{-32},  \label{dt}
\end{equation}
the Universe should be expanded $e^{100}\approx 10^{44}$ times,
i.e.
\begin{equation}
a(t_i+\triangle t)=10^{44}a(t_i).  \label{f}
\end{equation}
Therefore our complex scalar quintessence can also give an
explanation of the inflation in the early Universe.

\section{Conclusion}

\indent \hskip 16pt We present a new Lorentz gauge invariant
topological $ U(1)$ field theory. By virtue of the decomposition
theory of gauge potential and the $\phi $--mapping topological
current theory, we prove that from the viewpoint of spacetime
defects the set of zero points of $\phi ^A$
 $(A=1,2)$ create cosmic strings. When $\phi ^A\neq 0$ $(A=1,2)$, the complex scalar order parameter field which has the
property of negative pressure can reasonably play a role of
quintessence i.e. so-called dark energy. We also want to point out
that for the spacetime defects theory of cosmic strings it need
naturally introduce the order parameter field that just behave as
the quintessence of our Universe. We think that the further
investigation of creating quarks and leptons from cosmic strings
or quintessence is important.

\section*{Acknowledgments}

This work was supported by the National Natural Science Foundation
of China and the Doctoral Foundation of People's Republic of
China.

\end{document}